\documentclass{elsarticle}
\usepackage{graphicx}
\usepackage{xspace}

\usepackage{amsmath,amssymb}

\usepackage[dvips]{color}

\usepackage{latexsym,graphicx,epsfig}

\newcommand{\be}{\begin{equation}}
\newcommand{\ee}{\end{equation}}
\newcommand{\bea}{\begin{eqnarray}}
\newcommand{\eea}{\end{eqnarray}}
\newcommand{\bes}{\begin{subequations}}
\newcommand{\ees}{\end{subequations}}
\newcommand{\bear}{\begin{equation}\begin{array}}
\newcommand{\eear}[1]{\end{array}\label{#1}\end{equation}}
\newcommand{\fr}[2]{\dfrac{{ #1}}{{ #2}}}

\newcommand{\fn}[1]{\footnote{{#1}}}

\renewcommand{\le}{\leqslant}

\def\vep{{\varepsilon}}

\newcommand{\ggam}{\mbox{$\gamma\gamma\,$}}

{\end{list}}
\newcounter{enumct}

\begin{document}
\date{}
\title{{ What tell us LHC data about Higgs boson parity} }

\author{I.~F.~Ginzburg
\\
{\it Sobolev  Institute  of  Mathematics,  Department  of  Theor.  Physics,
Prosp. Ac. Koptyug, 4, and\\ Novosibirsk State University, ul. Pirogova,
2, 630090 Novosibirsk, Russia
}}

\begin{abstract}
Recently CMS and ATLAS announced that they had measured the  Higgs boson parity.  Here we note that their approach can determine this parity  only under the additional assumption  that this particle has a definite parity.

If parity conservation is violated in the Higgs sector, the parity of observed "Higgs" boson does not exist. The approach used in the mentioned experiments does not allow to observe  such opportunity. In this sense titles of mentioned CMS and ATLAS  publications are misleading.

\end{abstract}

\maketitle


In  the papers \cite{PexpCMS1}-\cite{PexpATLAS2}
the  CMS and ATLAS  collaborations declared that
they had measured the parity of $h(125)$ -- Higgs boson with mass of 125 GeV  (see also recent report \cite{Barr}).
Below we note that the titles of these papers\fn{"Study of the mass and spin-parity of the Higgs boson candidate via its decays to Z boson pairs" \cite{PexpCMS1}, "Constraints on the spin-parity and anomalous HVV
couplings of the Higgs boson in proton collisions at 7 and 8 TeV.
" \cite{PexpCMS2} and "Determination of spin and parity of the Higgs boson in the
$WW^* \to  e\nu\mu\nu$ decay channel with the ATLAS detector" \cite{PexpATLAS1}, "Study of the spin and parity of the Higgs boson in diboson decays with the ATLAS detector" \cite{PexpATLAS2}}
  are misleading.
The  approach used in \cite{PexpCMS1}-\cite{PexpATLAS2} can lead to a final result in the special case of the CP conserving Higgs sector only and does not work in the more general case.

To explain  our statement we review briefly  well known facts described e.g. in  \cite{Higgsrev1, Higgsrev2}.
The Higgs boson of SM is definitely P-even. The necessity of measuring its parity appears only in the extended models of the Higgs sector  -- (beyond Standard Model -- BSM).

(A) For the papers mentioned the method of measurement  was formulated in refs.~\cite{Ptheor1}-\cite{Ptheor3}. It is based on {\it the assumption that  the observed particle $h(125)$ has definite P-parity}.

If $h$ is P-even, it interacts with gauge bosons $W$ and $Z$  as in SM (for brevity we  write down only the interaction with $Z$,
the interaction with $W$ differs by coefficient only):
\be
\Delta L \propto h\cdot Z_\mu Z^\mu\,.\label{P+}
\ee

If $h$ is P-odd, its interaction to  gauge bosons comes from BSM interactions or from radiative corrections. The corresponding effective Lagrangian is given by   dimension 5 operator
\be
\Delta L\propto h\cdot\vep_{\mu\nu\alpha\beta} Z^{\alpha\beta} Z^{\mu\nu}\,.  \label{P-}
\ee

Papers \cite{Ptheor1}-\cite{Ptheor3} (see also \cite{Higgsrev1, Higgsrev2}) showed that
the forms \eqref{P+} and \eqref{P-} describe quite different correlations in the momentum distributions
of leptons produced in  decays $h\to ZZ^*\to (\ell_1\bar{\ell}_1)(\ell_2\bar{\ell}_2)$ or $h\to WW^*\to (\bar{\ell}_1\nu)(\ell_2\bar{\nu})$ with leptons $\ell=e,\,\mu$.

These correlations were measured in the experiments \cite{PexpCMS1}-\cite{PexpATLAS2} and it was found that correlations, correspondent to \eqref{P-} are absent (or small). This fact allows to conclude that these observations are consistent with the quantum numbers $J^{PC} = 0^{++}$ for $h(125)$.
\\

(B) However,   the dilemma discussed in \cite{Ptheor1}-\cite{Ptheor3}  is incomplete.
In many models for BSM physics, CP parity conservation in Higgs sector is violated, Higgs bosons have no definite P-parity. For such cases
{\bf the experiments \cite{PexpCMS1}-\cite{PexpATLAS2} are not
designed to observe this parity violation and therefore it is not possible
to draw any conclusions about $h(125)$ parity from these experiments}.

The common feature of BSM models is the presence of additional particles
similar to a Higgs boson -- both P-even and P-odd with their
possible mixing. We discuss for example the simplest example of such models -- the well known two-Higgs-doublet model, 2HDM (see, e.g., \cite{TDLee}) {\it(the Higgs sector of MSSM is its particular case).}

In this model the basic Higgs doublet $\phi_1$ is supplemented by a second scalar doublet $\phi_2$.
The interaction of Higgs boson with gauge bosons comes from
a kinetic term of the Lagrangian
$ D^\mu\phi_1^*D_\mu\phi_1+ D^\mu\phi_2^*D_\mu\phi_2$ where $D_\mu$ is covariant derivative, which includes fields $W_\mu$ and $Z_\mu$.

The Electroweak symmetry breaking   with standard decomposition for neutral components  $\phi_i^0=(v_ie^{i\xi_i}+\zeta_i+i\eta_i)/\sqrt{2}$ produces four neutral
fields  $\zeta_{1,2}$, \ $\eta_{1,2}$ (where $v_{1,2}$ are v.e.v.'s of the
fields $\phi_1$ and $|v_1|^2+|v_2|^2=v^2$, $v=246$~GeV). One  linear combination of
$\eta_{1,2}$ gives a neutral component of the Goldstone field $G^0$,
the orthogonal linear combination of $\eta_i$ is denoted by $\tilde{\eta}$.
In the CP-conserving case a linear combination of the fields $\zeta_i$  forms
two scalar Higgses $h$ and $H$, while  $\tilde{\eta}$  describes a
P-odd Higgs $A$.
In the CP-violating case the fields $\zeta_i$ and $\tilde{\eta}$ are mixed,
forming three Higgs fields $h_a$ that have no definite P-parity.
The interaction of $h_a$ with $Z$ comes from a kinetic term in precisely
the same way as in the SM and can be written as
\be
g^Z_{SM}\sum\limits_a \chi_a^Vh_aZ^\mu Z_\mu\,.\label{haZZ}
\ee

In this main approximation the  form of $h_aZZ$  interaction does not depend on the P-parity of $h_a$ (only the "P-even part of $h_a$" interacts with $Z$),  small terms  $\propto \vep_{\mu\nu\alpha\beta}Z^{\alpha\beta} Z^{\mu\nu}$ \eqref{P-} can appear only in radiative corrections. In this case results of  the experiments \cite{PexpCMS1}-\cite{PexpATLAS2} can not rule out admixture of P-odd component in the observed $h(125)$.

This parity non-conservation can be observed in the Yukawa interactions of Higgs boson. Unfortunately, there is no universal measure of parity non-conservation, which characterizes given boson $h(125)=h_1$. Indeed, the mixing angle of CP-odd component in $h_1$ and the phases of  $h_1\to\tau\bar{\tau}$ and $h_1t\bar{t}$ couplings are generally independent quantities. The methods for the study of CP violation in Higgs sector are collected in \cite{Higgsrev1,Higgsrev2}, at LHC the results may be obtained in  experiments $h_1\to\tau\bar{\tau}$ and $pp\to t\bar{t}\,h_1+...$.
\\

(C) {\it In order to finish our discussion, we show that -- within 2HDM -- even a big admixture of P-odd components in the observed Higgs boson does not contradict the modern data.} For this goal  we  show that,  with a suitable choice of parameters, the
same values of the  cross sections $gg\to h\to \ggam$ and
$gg\to h\to ZZ$ can be obtained both in the SM  and in the strongly CP-violating case of 2HDM with $h=h_1=h(125)$,
\bear{l}
\fr{\sigma(gg\to h_1\to \ggam)}{\sigma(gg\to h\to \ggam)_{SM}} =1\,,\qquad  \fr{\sigma(gg\to h_1\to ZZ)}{\sigma(gg\to h\to ZZ)_{SM}}= 1\,.
\eear{estim}
 Some  sets of  parameter values which satisfy these equations (with taking into account some other limitations) are presented in \cite{bigCP2HDM} for the cases when the $t\bar{t}h_1$ production cross section will differ significantly from its SM value.

We are interested in the case when, in addition to \eqref{estim}, future measurements of the $t\bar{t}h_1$ production cross section will give results which are  very close to predictions of the SM.

In our calculations
 we use relative couplings, determined
for the neutral Higgs bosons $h_a$ with mass $M_a$ and for the charged
Higgs bosons $H^\pm$ with mass $M_\pm$:
\bear{c}
\chi^P_{a}=\fr{g^P_a}{g^P_{\rm SM}} \;\; (P=V\,(W,Z) , q=(t,b,...))\,,\quad\chi^\pm_a=\fr{g(H^+H^-h_a)}{2M_\pm^2/v}\,.
\eear{relcoupldef}
(The ratio $\chi_3^V/\big(\chi_1^V\sqrt{1-(\chi_2^V)^2}\;\big)$ describes the admixture of the CP odd state in $h_1$  for 2HDM \cite{GKan15}.)

Using the well-known equations for the two-photon and
two-gluon widths, collected e.g. in \cite{TDLee}, \cite{Gintriple},
we determine two benchmark sets of  parameters, giving ratios \eqref{estim} at $|\chi_1^t|=1$:
\bear{c}
(I)\;\;\begin{array}{l} \chi_1^V=0.9,\quad \chi_1^\pm=0.4,
 \quad Re(\chi_1^t)=0.9, \;\;   Im(\chi_1^t)=0.43\,;\end{array}\\[3mm]
(II)\;\; \begin{array}{l}\chi_1^V=0.8, \quad \chi_1^\pm=1.4,\quad
 Re(\chi_1^t)=0.74,\;\; Im(\chi_1^t)=0.67\,.\end{array}
\eear{bench}
(We neglected all fermion contributions except $t$-quarks.)

The couplings $\chi_a^V$ obey a sum rule $\sum (\chi_a^V)^2=1$. Therefore,
in  case (I) the sum $(\chi_2^V)^2+(\chi_3^V)^2=0.19$, which allows us to have
$\chi_2^V\approx \chi_3^V\approx 0.3$ (the admixture of the P-odd to P-even
components of the $h_1$ about 0.3).
In  case (II) the sum $(\chi_2^V)^2+(\chi_3^V)^2=0.36$, which allows us to
have $\chi_2^V\approx \chi_3^V\approx 0.4$ (the admixture of the P-odd to
P-even components of the $h_1$ about 0.5).

This simple analysis shows that a big P-odd admixture in the observed Higgs boson is compatible with SM-like values for many observed quantities.

{\bf Summary}.   The  results \cite{PexpCMS1}-\cite{PexpATLAS2}  cannot  give  model independent information about  Higgs boson parity.
These results
allow to determine Higgs boson parity only under assumption that this particle has definite parity. In our opinion this limitation
 should be included in the titles \cite{PexpCMS1}-\cite{PexpATLAS2} for accuracy.
 \\

I am thankful S. Eidelman, B. Grzadkowski, I. Ivanov, J.Kalinowski, D.~Kazakov, M. Krawczyk,  G. Landsberg, K.~Melnikov, R. Santos and M.~Vysotsky for useful discussions.
This work was supported in part by the grants RFBR  15-02-05868,
NSh-3802.2012.2 and  NCN OPUS 2012/05/B/ST2/03306 (2012-2016).\\

\end{document}